\documentclass[12pt]{article}
\usepackage{amsmath}
\usepackage{amsthm}
\usepackage{graphicx,psfrag,epsf}
\usepackage{enumerate}
\usepackage{natbib}
\usepackage{amsfonts}
\usepackage{url} 
\usepackage{mathtools}
\newtheorem{theorem}{Theorem}[section]
\newtheorem{corollary}{Corollary}[theorem]

\newtheorem{definition}[theorem]{Definition}

\newcommand{\bb}{\boldsymbol}
\usepackage{graphicx}
\usepackage{multirow}
\DeclareMathOperator{\px}{\mathcal{P}_2(\mathcal{X})}
\newcounter{daggerfootnote}
\newcommand*{\daggerfootnote}[1]{%
    \setcounter{daggerfootnote}{\value{footnote}}%
    \renewcommand*{\thefootnote}{\fnsymbol{footnote}}%
    \footnote[2]{#1}%
    \setcounter{footnote}{\value{daggerfootnote}}%
    \renewcommand*{\thefootnote}{\arabic{footnote}}%
    }

\newcommand{\blind}{0}

\begin{document}

\def\spacingset#1{\renewcommand{\baselinestretch}%
{#1}\small\normalsize} \spacingset{1}


\if1\blind
{
  \title{\bf Distribution-in-distribution-out Regression}
  \author{Xiaoyu Chen, }\hspace{.2cm}\\
    Department of Industrial and Systems Engineering, University at Buffalo\\
    Mengfan Fu \\
    Department of Industrial and Systems Engineering, University at Buffalo
  \maketitle
} \fi

\if0\blind
{
  \bigskip
  \bigskip
  \bigskip
  \begin{center}
    {\Large \bf Distribution-in-distribution-out Regression}
    \vskip 5pt
    Xiaoyu Chen$^{a\dagger}$, Mengfan Fu$^{a}\protect\daggerfootnote{These authors contributed equally.}$, Yujing (Zipan) Huang$^{a}$, Xinwei Deng$^{b}$\footnote{Address for correspondence: Xinwei Deng, Professor, Department of Statistics, Virginia Tech, Blacksburg, VA 24061 (E-mail: xdeng@vt.edu).}
    \vskip 5pt
    {$^{a}$Department of Industrial and Systems Engineering, University at Buffalo, USA \\
    \vskip 3pt
    $^{b}$Department of Statistics, Virginia Tech, USA}

\end{center}
  \medskip
} \fi

\bigskip
\begin{abstract}

Regression analysis with probability measures as input predictors and output response has recently drawn great attention.
However, it is challenging to handle multiple input probability measures due to the non-flat Riemannian geometry of the Wasserstein space, hindering the definition of arithmetic operations, hence additive linear structure is not well-defined.
In this work, a distribution-in-distribution-out regression model is proposed by introducing parallel transport to achieve provable commutativity and additivity of newly defined arithmetic operations in Wasserstein space. 
The appealing properties of the DIDO regression model can serve a foundation for model estimation, prediction, and inference.
Specifically, the Fr\'echet least squares estimator is employed to obtain the best linear unbiased estimate, supported by 
the newly established Fr\'echet Gauss-Markov Theorem. Furthermore, we investigate a special case when predictors and response are all univariate Gaussian measures, leading to a simple close-form solution of linear model coefficients and $R^2$ metric.
A simulation study and real case study in intra-operative cardiac output prediction are performed to evaluate the performance of the proposed method.


\end{abstract}

\noindent%
\vfill

\newpage
\spacingset{1.45} 
\section{Introduction}
\label{sec:intro} 
Linear regression models have been widely used in many real applications to estimate the slope, representing the rate of change of output response variable ($Y-EY$) with respect to the change of input predictor variable ($X-EX$).
Benefit from flatness of Euclidean space (\citealt{bieberbach1911bewegungsgruppen}), the change in variables can be readily defined as line segments based on the Euclidean geometry, leading to appealing model estimation and inference.
However, the real-world measurements often encounter dependent and independent variables measured with uncertainty, represented as probability distributions.
Thus it poses significant challenges to the validity of using linear regression in Euclidean space for data with distributional inputs and outputs.
For example, an individual's non-invasive blood pressures can vary among different measurements.
Considering the uncertain measurements and individual differences, the clinical guidance and expected outcomes are usually provided as intervals (\citealt{practicalguideline}).
The conventional linear regression models cannot directly regress a dependent probability distribution on other independent probability distributions since they are in the non-Euclidean space.
It calls for a non-Euclidean regression model to better understand the relationship among uncertain measurements.

In general, many real applications are producing non-Euclidean data represented by probability distributions
For example, anesthesia machines controls the mixture of gases administered to a patient during surgery to prevent hypotension (low blood pressure) and/or hypertension (high blood pressure). 
However, both the control of the mixing ratios and the resulting blood pressures are generally non-Euclidean (i.e., they are either controlled or measured with uncertainties), owing to some unknown random processes. 
It is known that 
human beings have a homeostatic mechanisms of blood pressure auto-regulation, which adjusts blood pressure in real-time via Baroreceptor Reflex system and other factors (\citealt{johnson1986autoregulation, CHAPLEAU2012161}). 
The mixed effects of blood pressure auto-regulation and anesthesia gases are not directly measurable, making the blood pressure response an unknown process.
On the other hand, the input of the process are imperfectly controlled, which is normally quantified by control accuracy and precision. 
Therefore, both input and output can be are random processes, generating non-Euclidean data. 

One direct way to handle the uncertain measurements is to define input predictors and output response of a sample as probability measures that are sampled from the unknown processes. 
Assume that the probability measures are absolutely continuous w.r.t. the Lebesgue measure, 
we denote this type of data as the {\it distribution-in-distribution-out (DIDO) data}, 
and the regression model on such DIDO data as the {\it DIDO regression model}.
Modeling such DIDO data in a non-flat space raises significant challenges to existing statistical theories established in Euclidean space. 
One key challenge is that arithmetic operations, such as addition of probability measures and scalar multiplication with probability measures, are not well-defined on a non-flat space. For example, $p$-Wasserstein space as a set of probability measures was shown to be a Riemannian manifold with non-flat geometry (\citealt{ambrosio2005gradient}). 
For any two probability measures on the $p$-Wasserstein space, one cannot easily add them as in the Euclidean space since the result of such addition may not be a probability measure any more.
Another perspective to model the relationship among probability measures is to consider their densities as smooth functions, resulting in a function-on-function regression formulation (\citealt{sun2018optimal, morris2015functional}). 
However, constraining the functions within the space of probability density functions is non-trivial, which was realized by constrained functional approximation, see an example of Gaussian process-based approximation of nonparametric approximation \cite{tran2023learning} and a tensor product B-spline approximation in \cite{liu2023dynamic,guan2020estimating}. Such an approximation can be computationally expensive, and many approximation methods do not guarantee unit integral. 

Restricted by the nonexistence of arithmetic operations, regression models for DIDO data in the literature mainly consider the case with a single predictor. 
Notable methodologies include a distribution-on-distribution regression model (\citealt{ghodrati2022distribution}), which formulates the regression problem via optimal transport theory by learning an average map that transports the input probability measure to the response measure.
This map, as the learnable model coefficients, is averaged over all samples by minimizing a strictly convex Fr\'echet least squares objective function with monotonic constraints, which are estimated by the isotonic regression. 
Learning the average map depends on piece-wise constant approximation and often requires data of a large sample size.
Such univariate distribution-on-distribution regression model motivated the investigation of a multivariate Gaussian case (\citealt{okano2023distribution}), where the predictor and the response are all multivariate Gaussian measures. 
Their work constructs a nearly isometry map from Gaussian space to linear matrix (i.e., Euclidean) space such that the Gaussian-to-Gaussian regression model can be transformed to the linear matrix space with a certain inner product.
They hence defined the linear regression model with high dimensional tensor coefficients and investigated a low-rank approximation of the tensor.
By introducing tensor coefficients, this approach imposes high requirements on sample sizes, lacking direct interpretation of the original non-Euclidean data for practical inference.
\cite{chen2023wasserstein} identified another way to formulate the univariate regression problem in the Wasserstein space. 
A Wasserstein regression model was proposed by mapping the predictor and response probability measures onto the same point via parallel transport. The authors showed that the Wasserstein regression model connects to linear regression in Euclidean space. 
Motivated by this work, \cite{zhu2023geodesic} extended the Wasserstein regression model to multiple distribution-to-distribution regression where multiple predictors are presented by defining arithmetic operations on metric space of random objects. 
However, the newly defined addition and scalar multiplication do not enjoy commutativity, requiring extra optimization procedures to order predictors according to the variances they can explain. 

In this work, our focus is to enable multiple input predictors of probability measures for the DIDO regression model in Wasserstein space, which can have appealing properties for model estimation, prediction, and inference.
The key idea of the proposed DIDO regression model is to construct the relationship between the rate of change of independent probability measures and that of dependent probability measure.
Specifically,  we propose a DIDO regression model by defining commutable addition and scalar multiplication operations via parallel transport, hence enabling additivity in the Wasserstein space.
Benefit from the additivity and the strict convexity of the Fr\'echet least squares, the proposed method has several advantanges. First, the proposed DIDO regression model has a close-form optimal solution based on the Wasserstein covariance (\citealt{petersen2019wasserstein}). Second, we can establish the Gauss-Markov theorem for the proposed DIDO regression model, in the sense that the Fr\'echet least squares estimator yields the best linear unbiased estimator (BLUE). Third, the proposed DIDO regression has simple solution when predictors and response are all Gaussian measures, showing its connection to multiple linear regression in Euclidean space. 
Furthermore, we introduce the Dirac notation from quantum mechanics (\citealt{messiah2014quantum}) as a convenient tool to proof the Fr\'echet Gauss-Markov theorem, making the theoretical investigation of DIDO more appealing.

The proposed DIDO regression model generalizes linear regression from modeling Euclidean data to non-Euclidean measure-valued data. It is worth to remark that, different from the linear regression model in the Euclidean space, 
the DIDO regression model relaxes the \textit{i.i.d.} assumption by allowing non-identically distributed errors. Instead, the proposed DIDO regression assumes that the error probability measures are independently sampled from an identical process. 
Note that the proposed DIDO model maintains its BLUE property and the degree of freedoms, making it suitable for model estimation and inference needed in many real-world applications.

The reminder of this article is organized as follows. We first briefly review on Wasserstein space in Section~\ref{sec:prelim}, and then detail the proposed DIDO regression model based on newly defined arithmetic operations and their commutative and linear properties in Section~\ref{sec:dido}. 
In Section~\ref{sec:estimation}, we  investigate the Fr\'echet least squares estimator and establish the Fr\'echet Gauss-Markov theorem. 
To serve practical needs, a DIDO regression with Gaussian measures is developed in Section~\ref{sec:gaussian}. 
The DIDO model is then validated via a simulation study in Section~\ref{sec:simulation}, and a real case study of intra-operative cardiac output distribution prediction problem in Section~\ref{sec:case-study}. 
We conclude this work with discussions and future directions in Section~\ref{sec:conc}.

\section{Brief Review on Wasserstein Space}
\label{sec:prelim}


We first brief the foundation of optimal transport in Wasserstein space (\citealt{ghodrati2022distribution}). 
Given $\mathcal{X}$ as a separable metric space, we denote $\mathcal{P}(\mathcal{X})$ as the family of univariate probability measures on $\mathcal{X}$, and define the 2-Wasserstein distance between probability measures $\mu$ and $\nu$ as:
\begin{align*}
\label{eq:wasserstein_distance}
d^2_{W}(\nu, \mu) \coloneqq \inf_{\gamma \in \Gamma(\nu,\mu)} \int_{\Omega} |x - y|^2 d\gamma(x, y),
\end{align*}
where the infimum is taken over all couplings in $\Gamma(\nu, \mu)$ that take $\nu$ and $\mu$ as their marginals.
Equipping $\mathcal{P}(\mathcal{X})$ with a 2-Wasserstein distance, we further define the Wasserstein space of probability measures:
\begin{align*}
    \mathcal{P}_2(\mathcal{X}) \coloneqq \{\mu(x) \in \mathcal{P}(\mathcal{X}) : \exists x_0\in \mathcal{X} \text{ s.t. } E_{\mu}[d^2(x_0,x)]<+\infty \}.
\end{align*}
Note that the $\mathcal{P}_2(\mathcal{X})$ was proven to have Riemannian structure (i.e., a smooth manifold), enables the analysis of smooth curves, hence geodesics (\citealt{ambrosio2005gradient}).

We consider the case where a source probability measure, denoted by $\mu$, and a target probability measure, denoted by $\nu$, are absolutely continuous relative to Lebesgue measure, which guarantees that the optimal transport plan can be represented by a certain map $T$. Specifically, in the one-dimensional case, this mapping $T$ is uniquely determined by the composition of the quantile function of $\nu$ and $\mu$, denoted as ${F}^{-1}_{\nu}$ and ${F}^{-1}_{\mu}$. Consequently, the squared 2-Wasserstein distance between $\mu$ and $\nu$ can be quantified as:
\begin{equation}
\label{eq:wasserstein_distance}
d^2_W(\nu, \mu) = \int_0^1 (F^{-1}_{\mu}(t) - F^{-1}_{\nu}(t))^2 \, dt.
\end{equation}

To understand the rate of change of probability measures, it is essential to define extent of a path \( \gamma \) within any given metric space \( (X, d) \).
Given a path \( \gamma: [0, 1] \to X \), we define its arc length as:
\[
\text{Length}(\gamma) := \sup \left\{ \sum_{k=0}^{n-1} d(\gamma(t_k), \gamma(t_{k+1})) \mid n \geq 1, 0 = t_0 < t_1 < \ldots < t_n = 1 \right\}.
\]
This definition is not restricted to continuous paths; it includes any function from \( [0, 1] \) into \( X \). Paths with \( \text{Length}(\gamma) < \infty \) are precisely those of bounded variation, aligning with the concept of bounded variation (BV) functions. It is straightforward to ascertain that all paths \( \gamma \) belonging to the space of absolutely continuous functions AC(\(X\)) will have a finite arc length, i.e., \( \text{Length}(\gamma) \leq \int_0^1 g(t)dt < \infty \).
A differentiable curve parameterized by $\alpha\in [0,1]$ can be defined as \( \gamma(\alpha) : R \rightarrow \mathcal{P}_2(\mathcal{X}) \) with $\gamma(\alpha)\coloneqq (1-\alpha)\text{id}+\alpha T$, where id denotes the identity map. The trajectory is hence defined as $\mu_{\alpha}\coloneqq \gamma(\alpha)_\# \mu$, characterizing a constant-speed geodesic in $\mathcal{P}_2(\mathcal{X})$, which connects \( \mu_0 = \mu \) to \( \mu_1 = \nu \).

\section{Distribution-in-distribution-out Regression}
\label{sec:dido}

Regression models aim to learn a map from the input predictor(s) of the $i$-th sample $\mu_{i1}, \cdots, \mu_{ip}$ to transport the sample mean of output responses $\bar{\nu}_i$ to the observed response $\nu$. 
A parametric regression model of interest on $\mathcal{P}(\mathcal{X})$ is hence to find model parameters $\bb{\alpha}=(\alpha_1,\cdots,\alpha_q)^T$ that parameterizes a map $T_{0i}$, where $q$ is the number of model parameters, such that:
\begin{align}
    \nu_i &= (T_{\epsilon_i}\oplus T_{0i} )_\# \bar{\nu},
\end{align}
where $T_{\epsilon_i}$ is the random error map, and the addition operator $\oplus$ on $\mathcal{P}(\mathcal{X})$ will be defined in Section~\ref{sec:add}.
For a linear model, $T_{0i}$ is normally assumed to have linear structure reflecting the additive effects of predictors. However, such a linear model structure is non-trivial in $\mathcal{P}(\mathcal{X})$ since the space is no longer flat, where arithmetic operations are not well-defined. 
Therefore, we investigate how such a linear structure can be achieved in $\mathcal{P}(\mathcal{X})$.
In this section, we start from basic notations and general assumptions of the proposed DIDO regression model, then introduce the parallel transport and several basic arithmetic operations, which will be used to establish the foundation of the DIDO regression model.

\subsection{Wasserstein Covariance and Parallel Transport}

Given a univariate process $\mu$ with finite second moment which generates $N$ random probability measures $\{\mu_i\}_{i=1,...,N}$, jointly denoted as $\bb{\mu}$, the Fr\'echet barycenter and the Fr\'echet variance (\citealt{frechet1948two}) of $\mu$ are defined as:
\begin{align}
    \Bar{\mu} = \underset{\omega \in \px}{\arg\min} \frac{1}{N} \sum_{i=1}^N  d_W^2(\mu_i,\omega), \quad \text{var}(\bb{\mu}) = \frac{1}{N} \sum_{i=1}^N  d_W^2(\mu_i,\Bar{\mu}) . \label{eq:def-barycenter}
\end{align}

The Fr\'echet variance in (\ref{eq:def-barycenter}) can be further extended to Wasserstein-Fr\'echet (or Wasserstein for short) covariance based on a bivariate process $(\mu, \nu)$ each with finite second moment that generates $N$ random probability measures $\{\mu_i, \nu_i\}_{i=1,...,N}$ (\citealt{petersen2019wasserstein}).
Denote $F_{\Bar{\mu}}$ and $F_{\Bar{\nu}}$ as the law of Fr\'echet barycenters $\Bar{\mu}$ and $\Bar{\nu}$, respectively, the random optimal transport map from $\Bar{\mu}$ to $\mu_i$ is $T_{\mu_i,\Bar{\mu}} = F_{\mu_i}^{-1}\circ F_{\Bar{\mu}}$, similarly, $T_{\nu_i,\Bar{\nu}} = F_{\nu_i}^{-1}\circ F_{\Bar{\nu}}$, where $\circ$ denotes the functional composition.
To define the Wasserstein covariance in the same tangent space, the parallel transport map is hence defined to transport $T_{\mu,\Bar{\mu}}$ to the tangent space defined at $\Bar{\nu}$ by following the geodesic that connects $\Bar{\mu}$ and $\Bar{\nu}$ (\citealt{ambrosio2008construction}):
\begin{align} \label{eq:parallel}
    T_{\mu_i\|\Bar{\nu}} = T_{\mu_i,\Bar{\mu}} \circ T_{\Bar{\mu},\Bar{\nu}} - T_{\Bar{\mu},\Bar{\nu}} + \textup{id},
\end{align}
where $T_{\Bar{\mu},\Bar{\nu}}=F_{\Bar{\mu}}^{-1} \circ F_{\Bar{\nu}}$. 
Similarly, we have $T_{\nu_i\| \Bar{\mu}} = T_{\nu_i,\Bar{\nu}} \circ T_{\Bar{\mu},\Bar{\nu}}^{-1} - T_{\Bar{\mu},\Bar{\nu}}^{-1} + \textup{id}$, which is the transport map to transport $T_{\nu_i,\Bar{\nu}}$ to the tangent space defined at $\Bar{\mu}$. 
The sample Wasserstein covariance is hence defined based on the inner product in the tangent space at $\Bar{\mu}$:
\begin{align}
    \textup{cov}(\bb{\mu},\bb{\nu}) =  \frac{1}{N} \sum_{i=1}^N \left[ \int_{\mathcal{X}} (T_{\mu_i\|\Bar{\nu}}(x) - x)(T_{\nu_i,\Bar{\nu}}(x) - x)d(\Bar{\nu}(x)) \right] = \frac{1}{N} \sum_{i=1}^N \langle T_{\mu_i\|\Bar{\nu}}, T_{\nu_i,\Bar{\nu}} \rangle_{\Bar{\nu}}.
\end{align}
Such successful quantification of covariance structure between random probability measures, enabled by parallel transport, yields an important tool to establish the arithmetic operations hence the additive linear structure on $\mathcal{P}_2{(\mathcal{X})}$.

\subsection{Addition and Scalar Multiplication on $\px$}
\label{sec:add}
A DIDO regression model in the non-Euclidean space requires clear definition of addition operator and scalar
multiplication operator, which is not trivial on manifold $\px$. 
Different from \cite{zhu2023autoregressive,zhu2023geodesic}, we define the addition among two or more optimal transport maps and scalar multiplication based on parallel transport.
\begin{definition}[Addition between two optimal transport maps]
    Given two maps $T_{\mu_i,\Bar{\mu}} = F_{\mu_i}^{-1}\circ F_{\Bar{\mu}}$ and $T_{\nu_i,\Bar{\nu}} = F_{\nu_i}^{-1}\circ F_{\Bar{\nu}}$, the addition is defined as:
    \begin{align}
        T_{\mu_i,\Bar{\mu}} \oplus T_{\nu_i,\Bar{\nu}} \coloneqq T_{\mu_i\|\nu_i} \circ T_{\nu_i,\Bar{\nu}},
    \end{align}
    where $T_{\mu_i\|\nu_i}$ is the parallel transport that maps $T_{\mu_i,\Bar{\mu}}$ to the tangent space defined at $\nu_i$.
\end{definition}

The generalization of addition operator to more than two optimal transport maps can hence be defined, forming a natural definition of scalar multiplication $\alpha \odot T_{\mu_i,\Bar{\mu}}$ by decomposing $\alpha$ into its sign, integer, and decimal components. 

\begin{definition}[Scalar multiplication with an optimal transport map] \label{def:operators}
    $\forall \alpha \in R$, we first decompose $\alpha$ into three parts $\alpha = s(a+b)$: 
    \begin{enumerate}
        \item[(i) ] $b = \lfloor |\alpha| \rfloor$, indicating addition of $b$ components of geodesics $\gamma_{T_{\mu_i,\Bar{\mu}}^s}(1)$;
        \item[(ii) ] $a = |\alpha|-b$, indicating geodesic $\gamma_{T_{\mu_i,\Bar{\mu}}^s}(a)$;
        \item[(iii) ] $s = \textup{sign}(\alpha) \in \{ -1, 1 \}$; indicating the direction of geodesics.
    \end{enumerate}
    where $\gamma_{T_{\mu_i,\Bar{\mu}}^s}(a)=\textup{id} + a(T_{\mu_i,\Bar{\mu}}^{s}-\textup{id})$. 
    The scalar multiplication is hence defined as:
    \begin{align}
        \alpha \odot T_{\mu_i,\Bar{\mu}} &\coloneqq \gamma_{T_{\mu_i,\Bar{\mu}}^s}(a) \oplus \underbrace{ \gamma_{T_{\mu_i,\Bar{\mu}}^s}(1) \oplus \cdots \oplus \gamma_{T_{\mu_i,\Bar{\mu}}^s}(1)}_{\text{b components}}, \nonumber \\
                       &\coloneqq a \odot T_{\mu_i,\Bar{\mu}}^s \oplus \underbrace{ 1 \odot T_{\mu_i,\Bar{\mu}}^s \oplus \cdots \oplus 1 \odot T_{\mu_i,\Bar{\mu}}^s}_{\text{b components}}, \nonumber \\
                       &= a \odot T_{\mu_i,\Bar{\mu}}^s \oplus_{k=1}^{b}T_{\mu_i,\Bar{\mu}}^s.
    \end{align}
\end{definition}

\subsection{DIDO Regression Model}
Denote $\{\boldsymbol{\mu}_{i},\nu_i\}_{i=1,\cdots,N}$ as the observational data of N samples, where $\boldsymbol{\mu}_{i}$ is a $p$-vector of random probability measures with elements $\left( \mu_{i1},\cdots,\mu_{ij},\cdots,\mu_{ip} \right)^T$ representing $p$ predictors of the DIDO regression. 
Both $\mu_{ij}$ and $\nu_i$ are random measures in a metric space $\px$ endowed with a 2-Wasserstein distance. 
Without loss of generality, we further assume that the cumulative distribution functions (i.e., $F_{\mu_{ij}}$ and $F_{\nu_i}$) of such probability measures are known. 
Note that in practice, even a distribution function may not be observable, one can readily perform density estimation or adopt histogram based on samples to generate empirical estimates of the distribution function.

In analogy to the ordinary least squares regression in Euclidean space that assumes independently and identically distributed (\textit{i.i.d.}) unmodeled errors, the proposed DIDO regression model assumes \textit{i.i.d.} error measures, implying that $\textup{cov}((T_{\epsilon_i})_\# \nu,(T_{\epsilon_j})_\# \nu) =0$, $\forall i\neq j$, identity mean error map $E[T_{\epsilon_i} | \bb{\mu}_i]=\textup{id}$, and constant Fr\'echet variance $\textup{var}(T_{\epsilon_i}|\bb{\mu}_i)=\sigma^2$. 
In addition, we assume that the probability measures in consideration are absolutely continuous with respect to Lebesgue measure.

Following the definition of addition and scalar multiplication, we formulate DIDO regression model in Definition~\ref{def:dido}.
\begin{definition}[Distribution-in-distribution-out regression] \label{def:dido}
    \begin{align}
        \nu_i &= \left( T_{\epsilon_i} \oplus \alpha_1 \odot T_{\mu_{i1},\Bar{\mu}_1} \oplus \cdots \oplus \alpha_p \odot T_{\mu_{ip},\Bar{\mu}_p} \right)_\# \Bar{\nu}, \nonumber \\
              &= \left( T_{\epsilon_i} \oplus_{j=1}^p \alpha_j \odot T_{\mu_{ij},\Bar{\mu}_j} \right)_\# \Bar{\nu},\label{eq:dido}
    \end{align}
where $T_{\epsilon_i}$ is the noise map with its expected map $E[T_{\epsilon_i}|\bb{\mu}_i]=\textup{id}$ as the identity map, $T_{\mu_{ij},\Bar{\mu}_j}$ is the random optimal transport that maps the $i$-th realization of the $j$-th predictor $\mu_{ij}$ to the Fr\'echet barycenter of the $j$-th predictor $\Bar{\mu}_j$.
\end{definition} 

Different from multiple linear regression in the Euclidean space, where variables are commutative, it is not obvious whether the order of predictors will affect the DIDO regression model since the geodesics are defined on the non-flat manifold, where the commutativity on a non-flat manifold does not hold, i.e., $T_1\circ T_2\neq T_2\circ T_1$.
Intuitively, one should not worry about the order of input variables when fitting a linear regression model. 
We hence establish the theorem of commutativity with parallel transport maps:
\begin{theorem}[Commutativity when $0\leq\alpha\leq 1$]
\label{thm:commutable}
With multiple probability measures and $0\leq\alpha \leq 1$, two properties can be found:
\begin{enumerate}
    \item[(i)] $( \oplus_{j=1}^p T_{\mu_{ij},\Bar{\mu}_j} )_{\#} \Bar{\nu} = (\oplus_{j\in \textup{Perm}([p])} T_{\mu_{ij},\Bar{\mu}_j} )_{\#} \Bar{\nu}$, where $\textup{Perm}([p])$ denotes a random permutation of $p$ indices, which identifies the commutativity;
    \item[(ii)] $( \oplus_{j=1}^p \alpha_j \odot T_{\mu_{ij},\Bar{\mu}_j} )_{\#} \Bar{\nu} = ( \sum_{j=1}^p \alpha_j\left[(F_{\mu_{ij}}^{-1} - F_{\Bar{\mu}_j}^{-1})\circ F_{\Bar{\nu}} \right]+ \textup{id} )_{\#} \Bar{\nu}$.
\end{enumerate}
 
\end{theorem}

\begin{proof}
    Without loss of generality, we proof (i) by showing that the commutativity $(T_{\mu_{ij},\Bar{\mu}_j} \oplus T_{\mu_{ik},\Bar{\mu}_k})_\# \Bar{\nu} = (T_{\mu_{ik},\Bar{\mu}_k} \oplus T_{\mu_{ij},\Bar{\mu}_j})_\# \Bar{\nu}$ holds for all $j,k\in [p]$ and $j\neq k$.
    Starting from the left hand side, we first find the parallel transport $T_{\mu_{ik} \| \Bar{\nu}}$ that maps $T_{\mu_{ik},\Bar{\mu}_k}$ to the tangent space at $\Bar{\nu}$. 
    With $T_{\mu_{ik},\Bar{\mu}_k}=F_{\mu_{ik}}^{-1}\circ F_{\bar{\mu}_k}$, one can find:
    \begin{align*}
        T_{\mu_{ik} \| \Bar{\nu}} &= T_{\mu_{ik},\Bar{\mu}_k} \circ T_{\Bar{\mu}_k , \Bar{\nu}} - T_{\Bar{\mu}_k , \Bar{\nu}} + \textup{id}, \\
        &= F_{\mu_{ik}}^{-1}\circ F_{\bar{\mu}_k} \circ F_{\Bar{\mu}_k}^{-1} \circ F_{\Bar{\nu}} - F_{\Bar{\mu}_k}^{-1} \circ F_{\Bar{\nu}} + \textup{id}, \\
        &= (F_{\mu_{ik}}^{-1} - F_{\Bar{\mu}_k}^{-1}) \circ F_{\Bar{\nu}} + \textup{id}.
    \end{align*}
    Similarly, the parallel transport $T_{\mu_{ij} \| \Tilde{\Bar{\nu}}}$, where $\Tilde{\Bar{\nu}} = (T_{\mu_{ik} \| \Bar{\nu}})_\# \Bar{\nu}$, can be derived as:
    \begin{align*}
        T_{\mu_{ij} \| \Tilde{\Bar{\nu}}} &= T_{\mu_{ij},\Bar{\mu}_j} \circ T_{\Bar{\mu}_j , \Tilde{\Bar{\nu}}} - T_{\Bar{\mu}_j , \Tilde{\Bar{\nu}}} + \textup{id}, \\
        &= F_{\mu_{ij}}^{-1}\circ F_{\bar{\mu}_j} \circ F_{\Bar{\mu}_j}^{-1} \circ F_{\Tilde{\Bar{\nu}}} - F_{\Bar{\mu}_j}^{-1} \circ F_{\Tilde{\Bar{\nu}}} + \textup{id}, \\
        &= (F_{\mu_{ij}}^{-1} - F_{\Bar{\mu}_j}^{-1}) \circ F_{\Bar{\nu}} \circ T_{\mu_{ik} \| \Bar{\nu}}^{-1} + \textup{id}.
    \end{align*}
    Performing the addition:
    \begin{align*}
        T_{\mu_{ij},\Bar{\mu}_j} \oplus T_{\mu_{ik},\Bar{\mu}_k} &= (F_{\mu_{ij}}^{-1} - F_{\Bar{\mu}_j}^{-1}) \circ F_{\Bar{\nu}} \circ T_{\mu_{ik} \| \Bar{\nu}}^{-1} \circ T_{\mu_{ik} \| \Bar{\nu}} + T_{\mu_{ik} \| \Bar{\nu}}, \\
        &= (F_{\mu_{ij}}^{-1} - F_{\Bar{\mu}_j}^{-1}) \circ F_{\Bar{\nu}} + (F_{\mu_{ik}}^{-1} - F_{\Bar{\mu}_k}^{-1}) \circ F_{\Bar{\nu}} + \textup{id}.
    \end{align*}
    Similarly, $T_{\mu_{ij},\Bar{\mu}_j} \oplus T_{\mu_{ik},\Bar{\mu}_k} = (F_{\mu_{ij}}^{-1} - F_{\Bar{\mu}_j}^{-1}) \circ F_{\Bar{\nu}} + (F_{\mu_{ik}}^{-1} - F_{\Bar{\mu}_k}^{-1}) \circ F_{\Bar{\nu}} + \textup{id}$, which proves the commutativity (i).

    Proving (ii) is essentially to ensure that the bivariate scenario $\alpha_j \odot T_{\mu_{ij}, \bar{\mu}_j} \oplus \alpha_k \odot T_{\mu_{ik}, \bar{\mu}_k}$ holds. 
    Given $\alpha_j, \alpha_k \in [0,1]$ and the definition of geodesic in Definition~(\ref{def:operators}), it can be shown that:
    \begin{align*}
        T_{ \gamma_{ik}(\alpha_k)_\# \Bar{\mu}_k \| \bar{\nu}} = (\alpha_k \odot T_{\mu_{ik}, \bar{\mu}_k})_{ \gamma_{ik}(\alpha_k)_\# \Bar{\mu}_k \| \Bar{\nu}} &= [\textup{id} + \alpha_k( T_{\mu_{ik}, \bar{\mu}_k} - \textup{id} ) ] \circ T_{\Bar{\mu}_k \| \Bar{\nu}} - T_{\Bar{\mu}_k \| \Bar{\nu}} + \textup{id}, \\
          &= \alpha_k(T_{\mu_{ik},\Bar{\mu}_k} \circ T_{\Bar{\mu}_k \| \Bar{\nu}} - T_{\Bar{\mu}_k \| \Bar{\nu}}) + \textup{id},\\
          &= \alpha_k ((F_{\mu_{ik}}^{-1} - F_{\Bar{\mu}_k}^{-1}) \circ F_{\Bar{\nu}})  + \textup{id},
    \end{align*}
    where $\gamma_{ik}(\alpha_k)$ abbrevs $\gamma_{T_{\mu_{ik}, \bar{\mu}_k}}(\alpha_k)$, which is the geodesic parameterized by $\alpha_k$.
    Similarly, define $\Tilde{\Bar{\nu}} = (\alpha_k \odot T_{\mu_{ik}, \bar{\mu}_k})_\# \Bar{\nu}$, one can find:
    \begin{align*}
        T_{ \gamma_{ij}(\alpha_j)_\# \bar{\mu}_j \| \Tilde{\Bar{\nu}}} = (\alpha_j \odot T_{\mu_{ij}, \bar{\mu}_j})_{ \gamma_{ij}(\alpha_j)_\# \bar{\mu}_j \| \Tilde{\Bar{\nu}}} &= [\textup{id} + \alpha_j( T_{\mu_{ij}, \bar{\mu}_j} - \textup{id} ) ] \circ T_{ \gamma_{ik}(\alpha_k) \| \bar{\nu}} - T_{ \gamma_{ik}(\alpha_k) \| \bar{\nu}} + \textup{id}, \\
          &= \alpha_j(F_{\mu_{ij}}^{-1} - F_{\Bar{\mu}_j}^{-1}) \circ F_{\Bar{\nu}} \circ T_{ \gamma_{ik}(\alpha_k) \| \bar{\nu}}^{-1} + \textup{id}.
    \end{align*}
    Combining these results, one can find:
    \begin{align*}
        \alpha_j \odot T_{\mu_{ij}, \bar{\mu}_j} \oplus \alpha_k \odot T_{\mu_{ik}, \bar{\mu}_k} &= T_{ \gamma_{ij}(\alpha_j) \| \Tilde{\bar{\nu}}} \circ T_{ \gamma_{ik}(\alpha_k) \| \bar{\nu}}\\
          &= \alpha_j(F_{\mu_{ij}}^{-1} - F_{\Bar{\mu}_j}^{-1}) \circ F_{\Bar{\nu}} \circ T_{ \gamma_{ik}(\alpha_k) \| \bar{\nu}}^{-1} \circ T_{ \gamma_{ik}(\alpha_k) \| \bar{\nu}} + T_{ \gamma_{ik}(\alpha_k) \| \bar{\nu}}\\
          &= \alpha_j(F_{\mu_{ij}}^{-1} - F_{\Bar{\mu}_j}^{-1}) \circ F_{\Bar{\nu}} + \alpha_k(F_{\mu_{ik}}^{-1} - F_{\Bar{\mu}_k}^{-1}) \circ F_{\Bar{\nu}} + \textup{id}.
    \end{align*}
    One can use the same procedure to proof that $\alpha_k \odot T_{\mu_{ik}, \bar{\mu}_k} \oplus \alpha_j \odot T_{\mu_{ij}, \bar{\mu}_j}$ yields the same results. We hence conclude the proof of (ii).
\end{proof}

Theorem~(\ref{thm:commutable}) also implies the additivity of the addition operator in Wasserstein space $\px$:

\begin{theorem}[Additivity]
\label{thm:linearity}
$\forall \alpha\in R$ and $T_{\mu_{ij},\Bar{\mu}_j}=F_{\mu_{ij}}^{-1} \circ F_{\Bar{\mu}_j}$, with $|\alpha| = a + b$ and $s = \textup{sign}(\alpha)$, where $b = \lfloor |\alpha| \rfloor$ and $a = |\alpha|-b$, the following additivity holds: 
\begin{align*}
    ( a\odot T_{\mu_{ij},\Bar{\mu}_j}^{s} \oplus \underbrace{ T_{\mu_{ij},\Bar{\mu}_j}^{s} \oplus \cdots \oplus T_{\mu_{ij},\Bar{\mu}_j}^{s}}_{\text{b components}} )_{\#} \Bar{\nu} = ( \textup{id} + \alpha(T_{\mu_{ij} \| \Bar{\nu}} - \textup{id}) )_{\#} \Bar{\nu},
\end{align*}
where $T_{\mu_{ij} \| \Bar{\nu}}$ is the parallel transport map. 
\end{theorem}
\begin{proof}
    We first proof the case when $\alpha\geq 0$, i.e., $s=1$. Based on Theorem~(\ref{thm:commutable}, ii), setting $\bb{\alpha}=[a, \underbrace{1,\cdots, 1}_{b components}]^T$ and denoting $T_{\mu_{ij}, \Bar{\mu}_j} = T_{\mu_i, \Bar{\mu}}$, we have:
    \begin{align*}
        (\alpha \odot T_{\mu, \Bar{\mu}})_\# \bar{\nu} &= ( a\odot T_{\mu,\Bar{\mu}} \oplus \underbrace{ 1\odot T_{\mu,\Bar{\mu}} \oplus \cdots \oplus 1\odot T_{\mu,\Bar{\mu}}}_{\text{b components}} )_{\#} \Bar{\nu}\\
        &= (a(F_{\mu}^{-1} - F_{\Bar{\mu}}^{-1})\circ F_{\Bar{\nu}} + b(F_{\mu}^{-1} - F_{\Bar{\mu}}^{-1})\circ F_{\Bar{\nu}} + \textup{id})_\# \Bar{\nu} \\
        &= (\alpha(F_{\mu}^{-1} - F_{\Bar{\mu}}^{-1})\circ F_{\Bar{\nu}} + \textup{id})_{\#} \Bar{\nu} \\
        &= ( \textup{id} + \alpha(T_{\mu \| \Bar{\nu}} - \textup{id}) )_{\#} \Bar{\nu}
    \end{align*}    
    When $\alpha < 0$, i.e., $s=-1$, setting $|\bb{\alpha}|=[a, \underbrace{1,\cdots, 1}_{b components}]^T$, $T_{\mu_{ij}, \Bar{\mu}_j} = T_{\mu, \Bar{\mu}}$, and $T_{\mu, \Bar{\mu}}^{-1} = T_{\Bar{\mu},\mu}$, we have:
    \begin{align*}
        (\alpha \odot T_{\mu, \Bar{\mu}})_\# \bar{\nu} &= ( a\odot T_{\mu,\Bar{\mu}}^{-1} \oplus \underbrace{ 1\odot T_{\mu,\Bar{\mu}}^{-1} \oplus \cdots \oplus 1\odot T_{\mu,\Bar{\mu}}^{-1}}_{\text{b components}} )_{\#} \Bar{\nu}\\
        &= (a(F_{\Bar{\mu}}^{-1} - F_{\mu}^{-1})\circ F_{\Bar{\nu}} + b(F_{\Bar{\mu}}^{-1} - F_{\mu}^{-1})\circ F_{\Bar{\nu}} + \textup{id})_\# \Bar{\nu} \\
        &= (-|\alpha|(F_{\mu}^{-1} - F_{\Bar{\mu}}^{-1})\circ F_{\Bar{\nu}} + \textup{id})_{\#} \Bar{\nu} \\
        &= ( \textup{id} + \alpha(T_{\mu \| \Bar{\nu}} - \textup{id}) )_{\#} \Bar{\nu},
    \end{align*}
    which suggests the same results as the case when $\alpha\geq 0$, we hence conclude the proof.
\end{proof}

Combining Theorem~(\ref{thm:commutable}) and Theorem~(\ref{thm:linearity}), the DIDO model can be expressed in an explicit format as follows:
\begin{corollary}[Explicit Distribution-in-distribution-out regression] \label{cor:explicit-dido}
Define $ T_{i0} = \oplus_{j=1}^p \alpha_j \odot T_{\mu_{ij},\Bar{\mu}_j}$, the DIDO regression model in Definition~\ref{def:dido} can be reformulated as
\begin{align}
    \nu_i = ( T_{\epsilon_i} \oplus T_{i0}  )_\# \Bar{\nu}, 
    \text{     where } T_{i0}   = \textup{id} + \sum_{j=1}^p {\alpha_j(T_{\mu_{ij}\|\Bar{\nu}} - \textup{id})}.
\end{align}
\end{corollary}
Corollary~(\ref{cor:explicit-dido}) enables efficient estimation of DIDO regression model via a strictly convex optimization problem.

\section{Estimation via Fr\'echet Least Squares}
\label{sec:estimation}
To estimate the DIDO regression model coefficient vector $\bb{\alpha}$, we can consider to minimize the expected 2-Wasserstein distance between the true response $\nu$ and the predicted response $\hat{\nu}$:
\begin{align} \label{eq:frechetLS}
    \bb{\alpha} & \coloneqq \underset{\bb{\alpha}\in R^p}{\text{argmin}} \frac{1}{2}E\left[ d_W^2(\nu,\hat{\nu}) \right].
\end{align}

\begin{theorem} \label{thm:exact-solution}
    Under the Fr\'echet Least Squares in (\ref{eq:frechetLS}), the parameter estimation of the proposed DIDO regression model in Definition~(\ref{def:dido}) has a simple close form solution:
    \begin{align}
        \bb{\alpha} = \Sigma^{-1} C,
    \end{align}
    where covariance matrix $\Sigma$ has element $\Sigma_{kl}=\textup{cov} (\mu_k,\mu_l)$, and cross-covariance vector $C$ has element $C_k = \textup{cov}(\mu_k, \nu)$, where $\textup{cov} (\mu_k,\mu_l) = E \left[ \int_{\mathcal{X}} (T_{\mu_k \| \mu_{\oplus l}}(x)-x)(T_{\mu_l, \mu_{\oplus l}}(x)-x) d(\mu_{\oplus l})\right]$ is the Wasserstein covariance, and $\mu_{\oplus l}$ represents the Fr\'echet barycenter of random probability measure $\mu_l$.
\end{theorem}
\begin{proof}
To proof this result, we start from reformulating the least squares:
    \begin{align}
        L(\bb{\alpha}) = \frac{1}{2}E\left[ d_W^2(\nu,\hat{\nu}) \right] 
        &= \frac{1}{2}E \left[ { \int_{\mathcal{X}} \left(T_{0}(x)-T_{\nu,\Bar{\nu}} (x) \right)^2 d\Bar{\nu}(x) } \right] \nonumber \\
        &= \frac{1}{2}E \left[ { \int_{\mathcal{X}} \left(x + \sum_{j=1}^p {\alpha_j(T_{\mu_{j}\|\Bar{\nu}}(x) - x)} -T_{\nu,\Bar{\nu}} (x)\right)^2 d\Bar{\nu}(x) }\right] \label{eq:ls-loss}
    \end{align}
As Equation~(\ref{eq:ls-loss}) is a strictly convex function w.r.t. $\bb{\alpha}$, we further show its partial derivative w.r.t. $\alpha_{j^*}$:
    \begin{align}
        \frac{\partial L(\bb{\alpha})}{\partial \alpha_{j^*}} &= 
        E \left[ \int_{\mathcal{X}} \left[x + \sum_{j=1}^p {\alpha_j(T_{\mu_{j}\|\Bar{\nu}}(x) - x)} -T_{\nu,\Bar{\nu}} (x) \right](T_{\mu_{j^*}\|\Bar{\nu}}(x) - x) d\Bar{\nu}(x) \right] \nonumber \\
        &=  \sum_{j=1}^p E \left[ \int_{\mathcal{X}} \alpha_j (T_{\mu_{j}\|\Bar{\nu}}(x) - x) (T_{\mu_{j^*}\|\Bar{\nu}}(x) - x) d\Bar{\nu}(x) \right] \nonumber \\ 
        & - E \left[ \int_{\mathcal{X}} (T_{\nu,\Bar{\nu}} (x) - x)(T_{\mu_{j^*}\|\Bar{\nu}}(x) - x) d\Bar{\nu}(x) \right] \nonumber \\
        &= \sum_{j=1}^p \alpha_j \textup{cov}(\mu_{j}, \mu_{j^*}) - \textup{cov}(\mu_{j^*}, \nu) \label{eq:ls-derivative}
    \end{align}
Setting (\ref{eq:ls-derivative})$=0$, we can find: 
    \begin{align}
        \alpha_{j^*} = \frac{\textup{cov}(\mu_{j^*}, \nu) - \sum_{j\neq j^*} \alpha_j \textup{cov}(\mu_{j}, \mu_{j^*})}{\textup{cov}(\mu_{j^*}, \mu_{j^*})} = \frac{\textup{cov}(\mu_{j^*}, \nu) - \sum_{j\neq j^*} \alpha_j \textup{cov}(\mu_{j}, \mu_{j^*})}{\textup{var}(\mu_{j^*})}. \label{eq:alpha-j}
    \end{align}
Reorganizing Equation~(\ref{eq:alpha-j}) into matrix form using Cramer's rule (\citealt{cramer1750introduction}) we have $\bb{\alpha} = \Sigma^{-1} C$, which concludes the proof. 
\end{proof}

Theorem~(\ref{thm:exact-solution}) enables a new quadratic estimator that is equivalent to the Fr\'echet Least Squares estimator in (\ref{eq:frechetLS}):
\begin{corollary}[Quadratic estimator of DIDO regression model] \label{cor:quad-estimator}
The parameters in the proposed DIDO model can be efficiently estimated by solving the following unconstrained quadratic programming problem:
    \begin{align}
        \bb{\alpha} \coloneqq &\underset{\bb{\alpha} \in R^p}{\arg \min}  \frac{1}{2}\bb{\alpha}^T\Sigma^{-1}\bb{\alpha} - C^T\bb{\alpha}. \label{est:quad}
    \end{align}
\end{corollary}
One can easily show that the expression in~(\ref{est:quad}) provides the same unique solution discussed in Theorem~(\ref{thm:exact-solution}) by applying Karush-Kuhn-Tucker condition.
It is worth to remarking that Theorem~(\ref{cor:quad-estimator}) enables model regularization by constraining the space of $\bb{\alpha}$:
\begin{align*}
        \boldsymbol{\alpha} \coloneqq & \underset{\boldsymbol{\alpha} \in R^p}{\textup{argmin}} \frac{1}{2}\boldsymbol{\alpha}^T\Sigma^{-1}\boldsymbol{\alpha} - C^T\boldsymbol{\alpha}, \\
        & \text{s.t.   }  J(\bb{\alpha}) \leq t,
\end{align*}
where $J(\bb{\alpha})$ penalizes the model complexity; $t > 0$ is a tuning parameter that controls the level of penalization. 

\subsection{Fr\'echet Best Linear Unbiased Estimation}
\label{sec:unbiased}
In analogy of ordinary least squares in Euclidean space, we then investigate the property of the Fr\'echet least square estimator by understanding its bias and variance. 
With $N$ samples, the Fr\'echet least square estimator can be approximated by:
\begin{align}
        L_N(\bb{\alpha}) = 
        \frac{1}{2}E\left[ d_W^2(\nu,\hat{\nu}) \right] 
        &\approx 
        \frac{1}{2N}\sum_{i=1}^{N}{ d_W^2(\nu_i,\hat{\nu}_i) }, \nonumber
\end{align}
which results in the estimate $\hat{\bb{\alpha}}=\tilde{\Sigma}^{-1}\tilde{C}$, where the sample covariance matrix $\tilde{\Sigma}$ has element $\tilde{\Sigma}_{kl}=\textup{cov} (\bb{\mu}_k,\bb{\mu}_l)$, and the sample cross-covariance vector $\tilde{C}$ has element $\tilde{C}_k = \textup{cov}(\bb{\mu}_k, \bb{\nu})$. 
Here $\bb{\mu}_k = (\mu_{1k}, \cdots, \mu_{Nk})$, and $\bb{\nu} = (\nu_1, \cdots, \nu_N)$ are vectors of $N$ samples. 
The following result is to show that the Gauss-Markov Theorem holds for Fr\'echet least square estimator such that $E[\hat{\bb{\alpha}}]=\bb{\alpha}$. 
\begin{theorem}[Fr\'echet Gauss-Markov Theorem]
    Assuming the independence among random error maps $\textup{cov}((T_{\epsilon_i})_\# \nu,(T_{\epsilon_j})_\# \nu) =0$, $\forall i\neq j$, identity mean error map $E[T_{\epsilon_i} | \bb{\mu}_i]=\textup{id}$, and constant Fr\'echet variance $\textup{var}(T_{\epsilon_i}|\bb{\mu}_i)=\sigma^2$, the Fr\'echet least square estimator is the best linear unbiased estimator (BLUE).
\end{theorem}
\begin{proof}
    We first prove that the Fr\'echet least square estimator finds a unique optimal solution. 
    Based upon Equation~(\ref{eq:ls-derivative}), we can derive the second order derivative w.r.t. $\alpha_{j^*}$:
    \begin{align}
        \frac{\partial^2 L(\bb{\alpha})}{\partial \alpha_{j^*}^2} &= \textup{cov}(\mu_{j^*},\mu_{j^*}) = \textup{var}(\mu_{j^*}) \geq0.
    \end{align}
    Since $L(\bb{\alpha})$ is strictly convex w.r.t. $\alpha_{j^*}$, we conclude that the Fr\'echet least square estimator can provide an unique optimal solution.
    
    We then prove that the Fr\'echet least square estimate $\hat{\bb{\alpha}}=\tilde{\Sigma}^{-1}\tilde{C}$ is unbiased.
    Denote $\bb{\mu}$ as an element in the $N\times p$ product space of $\mathcal{P}_2(\mathcal{X})$, $\bb{\mu}_{i}=(\mu_{i1},\cdots,\mu_{ip})$ as the $i$-th sample, and $\bb{\mu}_{j}=(\mu_{1j},\cdots,\mu_{Nj})^T$ as the $j$-th predictor. We also denote $\tilde{C}$ as the predicted sample cross-covariance as a function of $\bb{\alpha}$.
    To support the proof, we first introduce the Dirac notation from quantum mechanics (\citealt{messiah2014quantum}) to ease the derivations in the tangent space at $\Bar{\nu}$.
    Given $T_1$, $T_2$, $T_3$ as three maps that are transported to $\bar{\nu}$, the inner product $\langle T_1, T_2\rangle_{\bar{\nu}}$ is separated as a bra $\langle T_1 |$, and a ket $| T_2 \rangle$ such that $\langle T_1, T_2\rangle_{\bar{\nu}} = \langle T_1 |\cdot | T_2 \rangle$. In addition, the Dirac notations enjoys the rules of linear algebra: $| aT_2 \rangle = a| T_2 \rangle$, $| T_1 + T_2 \rangle = | T_1 \rangle + | T_2 \rangle$, etc.
    
    We first prove that the conditional expectation w.r.t the noise maps $T_{\epsilon_i}$ holds: $E[\hat{\bb{\alpha}}|\bb{\mu}] = \bb{\alpha}$.
    Noticing that $E[\hat{\bb{\alpha}}|\bb{\mu}] = E[\tilde{\Sigma}^{-1}\Tilde{C} | \bb{\mu}]  = \tilde{\Sigma}^{-1} E[\Tilde{C} | \bb{\mu}]$, we investigate the $j$-th element of $\Tilde{C}$:
    \begin{align*}
        E[\Tilde{C}_j|\bb{\mu}] &= E\left[ \frac{1}{N} \sum_{i=1}^N  \langle T_{\mu_{{ij}}||\bar\nu}| \cdot |T_{\nu_{i},\bar\nu} \rangle |\bb{\mu} \right]\\
        &= \frac{1}{N} \sum_{i=1}^N [ \langle T_{\mu_{{ij}}||\bar\nu}| \cdot E[|T_{\nu_{i},\bar\nu}|\bb{\mu}\rangle]] \\
        &= \frac{1}{N} \sum_{i=1}^N [ \langle T_{\mu_{{ij}}||\bar\nu}| \cdot E[|T_{\epsilon_i} \oplus T_{i0}|\bb{\mu}\rangle]] \\
        &= \frac{1}{N} \sum_{i=1}^N  \langle T_{\mu_{{ij}}||\bar\nu}, T_{i0} \rangle_{\bar\nu} = \hat{C}_j,
    \end{align*}
    which indicates that $E[\hat{\bb{\alpha}}|\bb{\mu}]=\Tilde{\Sigma}^{-1}\hat{C}$. Corollary~(\ref{cor:explicit-dido}) gives an explicit expression for $T_{i0}$, resulting in a nice form of $\hat{C}$:
    \begin{align*}
        \hat{C}_j &= \frac{1}{N} \sum_{i=1}^N \langle T_{\mu_{ij}||\bar\nu} | \cdot | \textup{id} + \sum_{k=1}^p {\alpha_k(T_{\mu_{ik}\|\Bar{\nu}} - \textup{id})}\rangle_{\bar\nu}, \\
        &= \frac{1}{N} \sum_{i=1}^N \langle T_{{\mu_{ij}||\bar\nu}} | \cdot \left( |(1-\sum_{k=1}^p \alpha_k) \textup{id}\rangle + \sum_{k=1}^p \alpha_k | {T_{\mu_{ik}\|\Bar{\nu}}}\rangle \right), \\
        &= \sum_{k=1}^p \alpha_k \frac{1}{N} \sum_{i=1}^N \langle T_{{\mu_{ij}||\bar\nu}} , {T_{\mu_{ik}\|\Bar{\nu}}}\rangle_{\bar\nu}, \\
        &= \sum_{k=1}^p \alpha_k \textup{cov}(\bb{\mu}_{j},\bb{\mu}_{k}) = \textup{cov}(\bb{\mu}_j,\bb{\mu})\bb{\alpha},
    \end{align*}
    which indicates the vector form $\hat{C}=\textup{cov}(\bb{\mu},\bb{\mu})\bb{\alpha}=\Tilde{\Sigma}\bb{\alpha}$. We hence find that $E[\hat{\bb{\alpha}}|\bb{\mu}]=\Tilde{\Sigma}^{-1}\Tilde{\Sigma}\bb{\alpha} =\bb{\alpha}$. The Law of Iterated Expectations shows that $E[\hat{\bb{\alpha}}] = E_{\bb{\mu}}[E[\hat{\bb{\alpha}}|\bb{\mu}]]= \bb{\alpha}$. We hence proved that the Fr\'echet least squares is an unbiased estimator.

    In addition, we also investigate whether the Fr\'echet least squares yield the least variance among other linear estimators. To complete the proof, the matrix $\bb{T}_{\bb{\mu} || \bar{\nu}}$ is introduced, with its $(i,j)$-th elements denoted by $T_{{\mu_{ij}} || \bar{\nu}}$. Besides, we denote a vector $\bb{T}_{\bb\epsilon}$ with its $i$-th element $T_{\epsilon_i}$, which gives the assumption $\textup{var}[ \bb{T}_{\bb\epsilon}|\bb\mu] = \sigma^2 I$, $I$ is an identity matrix with dimension $N\times N$. We first prove that the conditional covariance matrix of the Fr\'echet least square estimate is $\textup{var}[\hat{\bb\alpha}|\bb{\mu}] = \sigma^{2} {\tilde{\Sigma}}^{-1}$.
   \begin{align*}
    \textup{var}[\hat{\bb\alpha}|\bb{\mu}] 
    &=  \textup{var}[(\langle \bb{T}_{\bb{\mu} || \bar{\nu}} | \cdot | \bb{T}_{\bb{\mu} || \bar{\nu}}  \rangle) ^{-1} \langle  \bb{T}_{\bb{\mu} || \bar{\nu}} | \cdot |  \bb{T}_{\bb{\epsilon} || \bar{\nu}} \rangle | \bb\mu ], \\
    &= [{\tilde{\Sigma}}^{-1} \cdot \langle  \bb{T}_{\bb{\mu} || \bar{\nu}}|]\textup{var}[ \bb{T}_{\bb\epsilon}|\bb\mu] [{\tilde{\Sigma}}^{-1} \cdot \langle  \bb{T}_{\bb{\mu} || \bar{\nu}}|]^{T}, \\
    &= [{\tilde{\Sigma}}^{-1} \cdot \langle  \bb{T}_{\bb{\mu} || \bar{\nu}} |] {\sigma^{2}}I [  |\bb{T}_{\bb{\mu} || \bar{\nu}} \rangle \cdot {\tilde{\Sigma}}^{-1}] = {\sigma^{2}}{\tilde{\Sigma}}^{-1}.
    \end{align*}
    Since we are considering the set of estimates, we can write any estimate in this set as $\tilde{\bb\alpha} = A \cdot |  \bb{T}_{\bb{\nu} || \bar{\nu}}  \rangle$, where $A$ is a $p \times N$ matrix. Then, we can define $D$ as the discrepancy:
    \begin{align}
        D = A - [{\tilde{\Sigma}}^{-1} \cdot \langle  \bb{T}_{\bb{\mu} || \bar{\nu}} |]. \label{eq:D}
    \end{align}
    Substituting $A$ by Equation~(\ref{eq:D}):
    \begin{align*}
        \tilde{\bb\alpha}
        &= A \cdot |\bb{T}_{\bb{\nu} ||\bar{\nu}}  \rangle, \\
        &= D \cdot |\bb{T}_{\bb{\nu} || \bar{\nu}} \rangle +  [{\tilde{\Sigma}}^{-1} (\langle  \bb{T}_{\bb{\mu} || \bar{\nu}} | \cdot | \bb{T}_{\bb{\nu} || \bar{\nu}}  \rangle)],\\
        &= D \cdot |\bb{T}_{\bb{\nu} || \bar{\nu}}  \rangle+ \hat{\bb\alpha}.
    \end{align*}
    Taking an expectation on both sides:
        \begin{align*}
            E[\tilde{\bb\alpha}|\bb{\mu}] 
            &= E[D \cdot |\bb{T}_{\bb{\nu} ||\bar{\nu}}  \rangle+ \hat{\bb\alpha} | \bb{\mu}],\\
            &= E[D \cdot |\bb{T}_{\bb{\nu} ||\bar{\nu}}  \rangle| \bb{\mu}] +E[ \hat{\bb\alpha} | \bb{\mu}], \\
            &= E[D \cdot |\bb{T}_{\bb{\nu} ||\bar{\nu}}  \rangle| \bb{\mu}] + \bb\alpha, \\
            &= E[D (|\bb{T}_{\hat{\bb{\nu}} ||\bar{\nu}}  \rangle + |\bb{T}_{{\epsilon} ||\bar{\nu}}  \rangle) |\bb{\mu}] + \bb\alpha,\\
            &= D \cdot |\bb{T}_{\hat{\bb{\nu}} ||\bar{\nu}}  \rangle + \bb\alpha,\\
            &= D \cdot |\bb{T}_{{\bb{\mu}} ||\bar{\nu}}  \rangle {\bb\alpha}+{\bb\alpha},
        \end{align*}
    which indicates that $E[\tilde{\bb\alpha}|\bb{\mu}]=\bb\alpha$ only if $ D \cdot |\bb{T}_{{\bb{\mu}} ||\bar{\nu}}\rangle = \bb{0}$.
    Therefore, we can also prove that 
    \begin{align*}
        \textup{var}[\tilde{\bb\alpha}|\bb{\mu}]
        &= \textup{var}[D \cdot |\bb{T}_{\bb{\nu} ||\bar{\nu}}  \rangle+ \hat{\bb\alpha} | \bb{\mu}] \\
        &= \textup{var}[D (|\bb{T}_{\hat{\bb{\nu}} ||\bar{\nu}}  \rangle + |\bb{T}_{\bb{\epsilon} ||\bar{\nu}}  \rangle) + {\tilde{\Sigma}}^{-1} (\langle  \bb{T}_{\bb{\mu} || \bar{\nu}}| \cdot |\bb{T}_{\bb{\nu} ||\bar{\nu}}  \rangle) | \bb\mu] \\
        &=  \textup{var}[D |\bb{T}_{\hat{\bb{\nu}} ||\bar{\nu}}  \rangle  + D|\bb{T}_{\bb{\epsilon} ||\bar{\nu}}  \rangle + {\tilde{\Sigma}}^{-1}(\langle  \bb{T}_{\bb{\mu} || \bar{\nu}}| \cdot (|\bb{T}_{\bb{\epsilon} ||\bar{\nu}}  \rangle + |\bb{T}_{\hat{\bb{\nu} }||\bar{\nu}}  \rangle) | \bb\mu] \\
        &= \textup{var} [(D +{\tilde{\Sigma}}^{-1} \cdot \langle  \bb{T}_{\bb{\mu} || \bar{\nu}}| ) \cdot |\bb{T}_{{\bb{\epsilon}} ||\bar{\nu}}  \rangle | \bb\mu] \\
        &= [D + {\tilde{\Sigma}}^{-1} \cdot \langle  \bb{T}_{\bb{\mu} || \bar{\nu}}|] \cdot \textup{var}[\bb{T}_{\bb\epsilon}|\bb\mu] \cdot [D + {\tilde{\Sigma}}^{-1} \cdot \langle  \bb{T}_{\bb{\mu} || \bar{\nu}}|]^{T} \\
        &= [D + {\tilde{\Sigma}}^{-1} \cdot \langle  \bb{T}_{\bb{\mu} || \bar{\nu}}|]{\sigma^{2}}I [D + {\tilde{\Sigma}}^{-1} \cdot \langle  \bb{T}_{\bb{\mu} || \bar{\nu}}|]^{T} \\
        &= {\sigma^{2}}DD^{T}+{\sigma^{2}}{\tilde{\Sigma}}^{-1} \\
        &= {\sigma^{2}}DD^{T}+\textup{var}[\hat{\bb\alpha}|\bb{\mu}]
    \end{align*}
    Consequently,  the difference in conditional variance, $\textup{var}[\tilde{\bb\alpha}|\bb{\mu}] - \textup{var}[\hat{\bb\alpha}|\bb{\mu}] = {\sigma^{2}}DD^{T}$ is positive semi-definite. This is holds for any unbiased linear estimate $\tilde{\bb\alpha}$. Therefore, we conclude the proof that Fr\'echet least squares estimator is BLUE.
    
\end{proof}

\section{DIDO Regression with Gaussian Measures}
\label{sec:gaussian}
In this section, we investigate a special case when both the predictors and responses are random Gaussian measures. Compared with free-shaped probability measures, the DIDO regression model under Gaussian setting (Gauss-DIDO) yields more efficient implementation, clearer interpretation, and it is expected to provide widely applicable tool for many scientific discovery and engineering modeling problems. 

We first denote $\mu_{i,j} \sim N(m_{ij}, \sigma_{ij}^2)$, $\nu_{i} \sim N(n_{i}, \eta_{i}^2)$ as the $i$-th sample of the predictor-response tuple $\{\bb{\mu}, \nu\}$. 
Before defining the explicit form of Gauss-DIDO regression model, we first derive some useful results in the following theorem.

\begin{theorem}[Close-form random optimal transport and parallel transport maps]\label{prop:gauss-maps}
    The following results provide the close-form random optimal transport and parallel transport maps:
    \begin{enumerate}
        \item[(i)] Given two random Gaussian measures $\mu \sim N(m, \sigma^2)$, $\nu \sim N(n, \eta^2)$, the optimal transport map $T_{\nu,\mu}(x) = n+\frac{\eta}{\sigma}(x-m)$, which satisfies $\nu = {T_{\nu,\mu}}_{\#} \mu$;
        \item[(ii)] Given a predictor-response tuple of the $i$-th sample $\{ \mu_i, \nu_i \}$ and their corresponding Fr\'echet barycenters ${\Bar{\mu},\Bar{\nu}}$, the parallel transport $T_{\mu_i \| \Bar{\nu}}(x)=(x-\Bar{n})\frac{\sigma_i-\Bar{\sigma}}{\Bar{\eta}} + x + (m_i - \Bar{m})$ maps the optimal transport $T_{\mu_i, \Bar{\mu}}$ to the tangent space defined at $\Bar{\nu}$;
    \end{enumerate}
\end{theorem}
\begin{proof}
    (i) can be directly derived as a special case of the optimal transport map between multivariate Gaussian measures (\citealt{peyre2019computational}, Remark 2.3.1). One can also proof it by using the fact that the optimal transport $T_{\nu,\mu}(x) = F_{\nu}^{-1}\circ F_{\mu}$ and using $F_{\nu}^{-1}(p)=n+\sqrt{2}\eta \textup{erf}^{-1}(2p-1)$, and $F_{\mu}(x)=\frac{1}{2}(1+\textup{erf}(\frac{x-m}{\sqrt{2}\sigma}))$, where $\textup{erf}(z)=\frac{2}{\sqrt{\pi}}\int_0^z e^{-t^2}dt$. 

    To derive (ii), we first use (i) to find:
    \begin{align*}
        T_{\mu_i, \Bar{\mu}}(x) &= m_i + \frac{\sigma_i}{\Bar{\sigma}}(x-\Bar{m}),\\
        T_{\Bar{\nu},\Bar{\mu}}(x) &= \Bar{m} + \frac{\Bar{\sigma}}{\Bar{\eta}}(x-\Bar{n}).
    \end{align*}
    Following Equation~(\ref{eq:parallel}), we can derive the parallel transport map as:
    \begin{align*}
        T_{\mu_i \| \Bar{\nu}}(x) &= (T_{\mu_i, \Bar{\mu}} \circ T_{\Bar{\nu},\Bar{\mu}})(x) - T_{\Bar{\nu}, \Bar{\mu}}(x) + x,\\
        &= m_i + \frac{\sigma_i}{\bar{\sigma}}\left(\bar{m} + \frac{\bar{\sigma}}{\bar{\eta}}(x - \bar{n})\right) - \frac{\sigma_i}{\bar{\sigma}} \bar{m} - \left(\bar{m} + \frac{\bar{\sigma}}{\bar{\eta}}(x - \bar{n})\right) + x,  \\
        &= (x-\Bar{n})\frac{\sigma_i-\Bar{\sigma}}{\Bar{\eta}} + x + (m_i - \Bar{m}).
    \end{align*}
    We hence conclude the proof.
\end{proof}

Theorem~(\ref{thm:commutable}) and (\ref{thm:linearity}) facilitate the derivation of the close form solution of the multiple Gauss-DIDO regression model.
Similar to simple Gauss-DIDO regression model, one can find:
\begin{align}
    \hat{\nu}_i &= N(\bar{n} + \sum_{j=1}^p \hat{\alpha}_j(m_{ij} - \bar{m}_j), (\bar{\eta} + \sum_{j=1}^p \hat{\alpha}_j(\sigma_{ij} - \bar{\sigma}_j))^2), \label{eq:predict-multiple-gauss-dido} \\
    \hat{\bb{\alpha}} &= \hat{\Sigma}^{-1}\hat{C},\label{eq:multiple-gauss-dido-alpha}
\end{align}
where the sample Wasserstein covariance matrix has element $\hat{\Sigma}_{jk}=\frac{1}{N} \sum_{i=1}^N [ (m_{ij} - \Bar{m}_j)(m_{ik} - \Bar{m}_k) + (\sigma_{ij} - \Bar{\sigma}_j)(\sigma_{ik} - \Bar{\sigma}_k) ]$, and the sample Wasserstein cross-variance matrix has element $\hat{C}_{j}=\frac{1}{N} \sum_{i=1}^N [ (m_{ij} - \Bar{m}_j)(n_i - \Bar{n}) + (\sigma_{ij} - \Bar{\sigma}_j)(\eta_i - \Bar{\eta}) ]$.

The residual transport map can also be found as:
\begin{align}
    T_{\epsilon_i}(x) &= n_i + \frac{\Bar{\eta}+\sum_{j=1}^p \hat{\alpha}_j(\sigma_{ij} -\Bar{\sigma}_j)}{\eta_i}[x-(\Bar{n}+ \sum_{j=1}^p\hat{\alpha}_j(m_{ij}-\Bar{m}_j))].
\end{align}

With Gaussian predictors and responses, the explained Fr\'echet sum of squares v.s. total Fr\'echet sum of squares can be explicitly computed to evaluate the goodness of fit via $R^2$:
\begin{align}
    R^2 &\coloneqq 1- \frac{\sum_{i=1}^N( d_W^2(\nu_i,\hat{\nu}_i) )}{\sum_{i=1}^N( d_W^2(\nu_i,\Bar{\nu}_i))} \nonumber \\
    &= 1- \frac{ \sum_{i=1}^N \left[ \left(\Bar{n}+\sum_{j=1}^p{\hat{\alpha}_j (m_{ij}-\Bar{m}_i)}  - n_i \right)^2 + \left(\Bar{\eta}+\sum_{j=1}^p{\hat{\alpha}_j (\sigma_{ij}-\Bar{\sigma}_i)}  - \eta_i \right)^2 \right]^2 }{\sum_{i=1}^N \left[ \left(\Bar{n} - n_i \right)^2 + \left(\Bar{\eta} - \eta_i \right)^2\right]}. \label{eq:r-square}
\end{align}

The Equivalence between the Gauss-DIDO regression model in $\mathcal{P}_2(\mathcal{X})$ and the linear regression model in Euclidean space can be established by assuming that the Gaussian measures share the same constant variance. i.e., $\sigma_{ij}^2=\Bar{\sigma}_j^2=\sigma_{j}^2$, and $\eta_i^2 = \Bar{\eta}^2=\eta^2$. 
This way, Equation~(\ref{eq:predict-multiple-gauss-dido}) reduces to:
\begin{align}
    \hat{n}_i &= \Bar{n} + \sum_{j=1}^p \hat{\alpha}_j(m_{ij}-\Bar{m}_j)
    = \sum_{j=1}^p \hat{\alpha}_j m_{ij} + \underbrace{(\Bar{n} - \sum_{j=1}^p \hat{\alpha}_j\Bar{m}_j)}_{\text{intercept}}.
\end{align}

Consequently, the sample Wasserstein covariance matrix and cross-covariance vector reduce to their Euclidean definitions:
\begin{align}
    \hat{\Sigma}_{jk}&=\frac{1}{N} \sum_{i=1}^N [ (m_{ij} - \Bar{m}_j)(m_{ik} - \Bar{m}_k)  ],\\
    \hat{C}_{j}&=\frac{1}{N} \sum_{i=1}^N [ (m_{ij} - \Bar{m}_j)(n_i - \Bar{n}) ].
\end{align}
Similarly, one can evaluate that the residual map reduces to the residual of linear regression:
\begin{align}
    T_{\epsilon_i}^{LR}(x) &= n_i + [x-(\Bar{n}+ \sum_{j=1}^p\hat{\alpha}_j(m_{ij}-\Bar{m}_j))].
\end{align}

Implementing the DIDO regression model with free-shape probability measures is computationally expensive since one need to approximate the transport maps $T_{\mu_{ij} \| \bar{\mu}_j}$ using non-decreasing numerical vectors. 
Such approximation is even harder to be implemented when predictor measures and response measures are with different supports, because the operations among vectorized $T_{\mu_{ij} \| \bar{\mu}_j}$ require one to expand the supports to the union of all supports, making the approximation inefficient in reality.
The proposed Gauss-DIDO regression model hence addresses the computational issue by providing an efficient solution that only depends on the parameters of the Gaussian measures. 
It also provides guidance for practitioners to easily prepare data to perform DIDO regression analysis.

\section{Simulation Study}
\label{sec:simulation}

We first evaluate the performance of Gauss-DIDO regression model in a simulation study. Simulation scenarios are generated based on the following steps:
\begin{enumerate}
    \item[1.] Define number of samples $N\in \{ 50, 100, 500 \}$, number of predictor measures $p \in \{ 2, 5, 10 \}$, and noise level $\zeta \in \{ 0.01, 0.1, 0.5, 1\}$.
    \item[2.] Generate random Wasserstein barycenters $\bar{\mu}_j = N(m_j, \sigma_j^2)$, $\bar{\nu} = N(n, \eta^2)$, where random variables $m_j, n \sim N(0,1)$, and $\sigma_j, \eta \sim \text{Uniform}(1,2)$.
    \item[3.] Generate distributions of random errors $\epsilon_i = N(s_i,\tau_i^2)$, where $s_i \sim N(0,\zeta^2)$, and $\tau_i \sim \text{Uniform}(0.01, \zeta)$.
    \item[4.] Generate linear model parameters $[\alpha_1,\cdots, \alpha_p]$, where $\alpha_j \in \text{Uniform}(-2,2)$.
    \item[5.] Generate $N$ samples $\{ [\mu_{i1},\cdots,\mu_{ip}],\nu_i \}_{i=1,\cdots,N}$ by $\mu_{ij} = N(m_j+\Delta m, (\sigma_j+\Delta \sigma)^2)$, $\hat{\nu}_i = \left( T_{\epsilon_i} \oplus_{j=1}^p \alpha_j \odot T_{\mu_{ij},\Bar{\mu}_j} \right)_\# \Bar{\nu}$, and randomly split them into 70\% training samples and 30\% testing samples.
\end{enumerate}

The proposed Gauss-DIDO model was then estimated in 1000 replications of the aforementioned procedure.
To evaluate the performance of the Gauss-DIDO regression model, in addition to $R^2$ defined by Equation~(\ref{eq:r-square}), we define the coefficient estimation error (CEE) to evaluate the estimation accuracy of model coefficients, and the testing mean squared error (MSE):
\begin{align*}
    \text{CEE} &= \sqrt{ \frac{\| \bb{\alpha} - \bb{\hat \alpha} \|_2^2}{p} }, \quad
    \text{MSE} = \frac{1}{N}\sum_{i=1}^{N}{ d_W^2(\nu_i,\hat{\nu}_i) }.
\end{align*}

In addition, we include a benchmark model based on linear regression (LR) model in Euclidean space. 
Since LR model cannot directly model probability measures, we consider to separately model the mean and log-variance of the response Gaussian measures by using two LR models. Specifically, to model the mean, we define the predictor-response tuple as $\{(m_{i1},\cdots,m_{i_p},\sigma_{i1},\cdots,\sigma_{i_p})^T, n_i\}_{i=1,\cdots,N}$. Similarly, to model the log-variance, we define the predictor-response tuple as $\{(m_{i1},\cdots,m_{i_p},\sigma_{i1},\cdots,\sigma_{i_p})^T, \eta_i\}_{i=1,\cdots,N}$.
With $p$ Gaussian predictor measures, the total number of LR model coefficients is $4p+2$. Comparing with the $p$ coefficients of the proposed Gauss-DIDO, LR model requires more degree of freedom for estimation.
The performance of the LR model is then evaluated based on the predicted Gaussian distribution using $R^2$ and the MSE. 

\begin{table}[p]
\caption{Summary of simulation results}
\label{tab:simulation}
\resizebox{\textwidth}{!}{%
\begin{tabular}{cccccccccc}
\hline
\multicolumn{1}{l}{}     & \multicolumn{1}{l}{} & \multicolumn{4}{c}{$\epsilon=0.01$}                            & \multicolumn{4}{c}{$\epsilon=0.1$}                            \\ \cline{3-10} 
\multicolumn{1}{l}{}     & \multicolumn{1}{l}{} & \multicolumn{2}{c}{DIDO}      & \multicolumn{2}{c}{LR}         & \multicolumn{2}{c}{DIDO}      & \multicolumn{2}{c}{LR}        \\ \cline{3-10} 
                         &                      & $R^2$         & Test MSE      & $R^2$         & Test MSE       & $R^2$         & Test MSE      & $R^2$         & Test MSE      \\ \hline
\multirow{3}{*}{$n=50$}  & $p=2$                & 0.999 (0.002) & 0.040 (0.042) & 0.930 (0.204) & 0.038 (0.083)  & 0.938 (0.117) & 0.046 (0.049) & 0.907 (0.114) & 0.038 (0.062) \\ \cline{2-10} 
                         & $p=5$                & 1.000 (0.000) & 0.085 (0.078) & 0.797 (0.270) & 0.505 (1.083)  & 0.990 (0.006) & 0.124 (0.133) & 0.799 (0.296) & 0.381 (0.971) \\ \cline{2-10} 
                         & $p=10$               & 1.000 (0.000) & 0.191 (0.197) & 0.791 (0.161) & 5.465 (31.570) & 0.996 (0.002) & 0.260 (0.258) & 0.753 (0.172) & 1.204 (1.976) \\ \hline
\multirow{3}{*}{$n=100$} & $p=2$                & 0.999 (0.001) & 0.021 (0.021) & 0.947 (0.095) & 0.073 (0.265)  & 0.953 (0.088) & 0.028 (0.026) & 0.920 (0.088) & 0.036 (0.079) \\ \cline{2-10} 
                         & $p=5$                & 1.000 (0.000) & 0.055 (0.064) & 0.815 (0.186) & 0.361 (0.765)  & 0.987 (0.010) & 0.045 (0.042) & 0.809 (0.211) & 0.345 (0.658) \\ \cline{2-10} 
                         & $p=10$               & 1.000 (0.000) & 0.101 (0.127) & 0.740 (0.256) & 0.989 (1.811)  & 0.995 (0.003) & 0.111 (0.118) & 0.777 (0.134) & 0.816 (1.135) \\ \hline
\multirow{3}{*}{$n=500$} & $p=2$                & 0.999 (0.005) & 0.003 (0.006) & 0.950 (0.094) & 0.031 (0.077)  & 0.950 (0.102) & 0.014 (0.004) & 0.904 (0.111) & 0.044 (0.065) \\ \cline{2-10} 
                         & $p=5$                & 1.000 (0.000) & 0.011 (0.013) & 0.809 (0.109) & 0.246 (0.179)  & 0.989 (0.007) & 0.023 (0.012) & 0.824 (0.107) & 0.271 (0.305) \\ \cline{2-10} 
                         & $p=10$               & 1.000 (0.000) & 0.018 (0.017) & 0.793 (0.088) & 0.492 (0.364)  & 0.995 (0.002) & 0.030 (0.018) & 0.765 (0.115) & 0.604 (0.635) \\ \hline
\multicolumn{1}{l}{}     & \multicolumn{1}{l}{} & \multicolumn{4}{c}{$\epsilon=0.5$}                             & \multicolumn{4}{c}{$\epsilon=1$}                              \\ \cline{3-10} 
\multicolumn{1}{l}{}     & \multicolumn{1}{l}{} & \multicolumn{2}{c}{DIDO}      & \multicolumn{2}{c}{LR}         & \multicolumn{2}{c}{DIDO}      & \multicolumn{2}{c}{LR}        \\ \cline{3-10} 
\multicolumn{1}{l}{}     & \multicolumn{1}{l}{} & $R^2$         & Test MSE      & $R^2$         & Test MSE       & $R^2$         & Test MSE      & $R^2$         & Test MSE      \\ \hline
\multirow{3}{*}{$n=50$}  & $p=2$                & 0.563 (0.206) & 0.324 (0.122) & 0.554 (0.194) & 0.316 (0.154)  & 0.268 (0.137) & 1.183 (0.355) & 0.280 (0.131) & 1.204 (0.372) \\ \cline{2-10} 
                         & $p=5$                & 0.794 (0.100) & 0.389 (0.153) & 0.701 (0.149) & 0.686 (0.996)  & 0.509 (0.130) & 1.280 (0.444) & 0.484 (0.120) & 1.495 (0.708) \\ \cline{2-10} 
                         & $p=10$               & 0.901 (0.035) & 0.510 (0.279) & 0.609 (0.827) & 2.971 (8.868)  & 0.690 (0.086) & 1.564 (0.507) & 0.592 (0.197) & 2.977 (2.241) \\ \hline
\multirow{3}{*}{$n=100$} & $p=2$                & 0.556 (0.192) & 0.296 (0.072) & 0.542 (0.178) & 0.302 (0.072)  & 0.274 (0.148) & 1.148 (0.283) & 0.275 (0.139) & 1.156 (0.285) \\ \cline{2-10} 
                         & $p=5$                & 0.789 (0.086) & 0.348 (0.096) & 0.664 (0.115) & 0.559 (0.477)  & 0.521 (0.110) & 1.249 (0.266) & 0.455 (0.112) & 1.524 (0.587) \\ \cline{2-10} 
                         & $p=10$               & 0.894 (0.038) & 0.410 (0.152) & 0.693 (0.221) & 1.363 (2.003)  & 0.676 (0.081) & 1.302 (0.345) & 0.539 (0.200) & 1.998 (1.456) \\ \hline
\multirow{3}{*}{$n=500$} & $p=2$                & 0.547 (0.190) & 0.278 (0.034) & 0.523 (0.171) & 0.300 (0.069)  & 0.268 (0.140) & 1.072 (0.112) & 0.260 (0.131) & 1.087 (0.115) \\ \cline{2-10} 
                         & $p=5$                & 0.785 (0.068) & 0.282 (0.035) & 0.674 (0.068) & 0.451 (0.153)  & 0.493 (0.115) & 1.110 (0.127) & 0.438 (0.093) & 1.275 (0.221) \\ \cline{2-10} 
                         & $p=10$               & 0.886 (0.029) & 0.290 (0.033) & 0.694 (0.079) & 0.833 (0.455)  & 0.648 (0.074) & 1.100 (0.111) & 0.522 (0.065) & 1.575 (0.500) \\ \hline
\end{tabular}
}
\end{table}

\begin{table}[]
\caption{Summary of DIDO parameter estimation errors}
\label{tab:simulation-pe}
\scriptsize
\centering
\begin{tabular}{cccccc}
\hline
\multicolumn{1}{l}{}     & \multicolumn{1}{l}{} & $\epsilon=0.01$ & $\epsilon=0.1$ & $\epsilon=0.5$ & $\epsilon=1$  \\ \hline
\multirow{3}{*}{$n=50$}  & $p=2$                & 0.002 (0.001)   & 0.026 (0.015)  & 0.140 (0.070)  & 0.284 (0.152) \\ \cline{2-6} 
                         & $p=5$                & 0.003 (0.001)   & 0.029 (0.010)  & 0.140 (0.058)  & 0.283 (0.088) \\ \cline{2-6} 
                         & $p=10$               & 0.003 (0.001)   & 0.033 (0.008)  & 0.161 (0.038)  & 0.320 (0.081) \\ \hline
\multirow{3}{*}{$n=100$} & $p=2$                & 0.002 (0.001)   & 0.020 (0.010)  & 0.100 (0.055)  & 0.182 (0.094) \\ \cline{2-6} 
                         & $p=5$                & 0.002 (0.001)   & 0.020 (0.006)  & 0.104 (0.035)  & 0.207 (0.070) \\ \cline{2-6} 
                         & $p=10$               & 0.002 (0.001)   & 0.022 (0.005)  & 0.110 (0.027)  & 0.218 (0.050) \\ \hline
\multirow{3}{*}{$n=500$} & $p=2$                & 0.001 (0.000)   & 0.009 (0.005)  & 0.043 (0.024)  & 0.086 (0.040) \\ \cline{2-6} 
                         & $p=5$                & 0.001 (0.000)   & 0.009 (0.003)  & 0.046 (0.015)  & 0.090 (0.030) \\ \cline{2-6} 
                         & $p=10$               & 0.001 (0.000)   & 0.009 (0.002)  & 0.045 (0.011)  & 0.097 (0.023) \\ \hline
\end{tabular}
\end{table}
The results of the mean values and standard deviations of the $R^2$ and test MSE are summarized in Table~\ref{tab:simulation}, and the results of the mean values and standard deviations of CEE errors are summarized in Table~\ref{tab:simulation-pe}.
It can be observed that the Gauss-DIDO model accurately recovered the true model coefficients in the scenario with low noise level $\zeta=0.01$.
The performance is found to reduce along with the increasing random noises.
Additionally, with increasing sample sizes, the Gauss-DIDO model achieves increasing $R^2$ and decreasing test MSE; with increasing number of predictors, the Gauss-DIDO model presents decreasing $R^2$ and increasing test MSE, due to the decreasing degree of freedom.
One interesting finding is that $R^2$ increases as more predictors were added while fixed the sample size and noise level. This is aligned with the $R^2$ metric of linear regression in Euclidean space, suggesting one to consider the adjusted-$R^2$: $R_{adj}^2 \coloneqq 1-\frac{(1-R^2)(N-1)}{N-p-1}$.

\section{Real Case Study}
\label{sec:case-study}

In perioperarive medicine, accurate quantification of cardiac output (CO) yields an important vital sign indicating the cardiac functions. In this section, a real case study in intraoperative cardiac output management is used to examine the performance of the proposed DIDO method. 
There are several invasive techniques to directly measure cardiac output. For example, pulmonary artery catheter (PAC) requires insertion into a major blood vessel, which carries risks of infection, bleeding, and thrombosis (\citealt{matthay1988bedside}). The risks of such invasive techniques triggered the development of non-invasive cardiac output measurement methods, mostly relying on regression models that predict cardiac output based on physiological signals (e.g., arterial blood pressures) (\citealt{slagt2014systematic}). 
These non-invasive techniques are usually constructed based on point-to-point prediction, which defines one sample as the pair of physiological signals and the cardiac output measured at one timestamp by assuming that samples are identically and normally distributed. 
However, in real practice, such samples are not identically distributed since physiological measurements are non-stationary and noisy.
Moreover, different patients' health conditions and different surgery types can lead to very different data distributions.
Thus it is not appropriate to assume that the response has constant variance, 
We demonstrate the effectiveness of the proposed Gauss-DIDO model by modeling the predictors and responses as probability measures.

\begin{figure}[h!]
    \centering
    \includegraphics[width=0.5\textwidth]{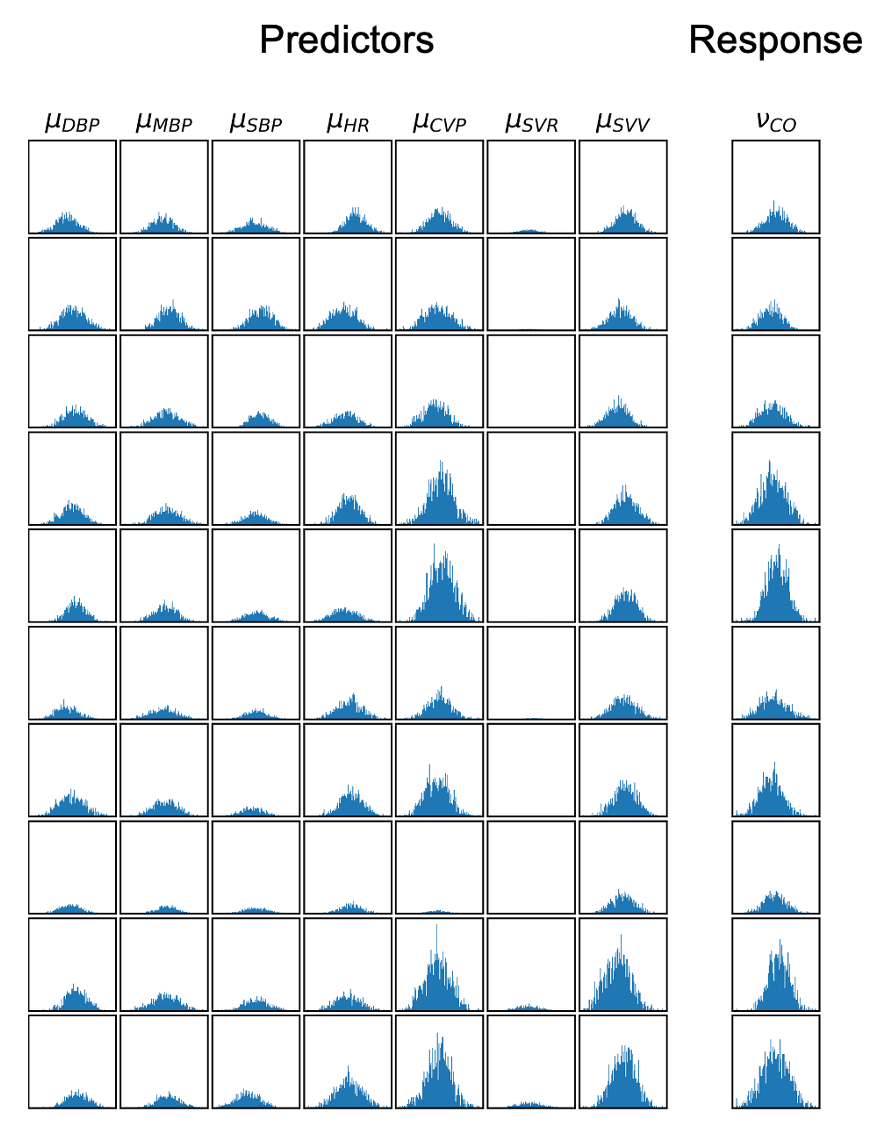}
    \caption{Distributional predictors and responses from intraoperative anesthesia records.}
    \label{fig:case-study-demo}
\end{figure}

Physiological signals from in total 232 surgery cases (24 surgery types) were collected from an open database referred as VitalDB (\citealt{lee2022vitaldb}), which is the largest database that stores vitals from both pre-, intra-, and post-operative phases of surgeries. We are particularly interested in understanding the relationship between seven predictor distributions (i.e., distributions of diastolic blood pressure (DBP), systolic blood pressure (SBP), mean blood pressure (MBP), hart rate (HR), central venous pressure (CVP), systemic vascular resistance (SVR), and stroke volume variation (SVV)) and one response distribution (i.e., cardiac output (CO)). 
Therefore, for each physiological signal and cardiac output in each surgery case, we firstly segmented the raw waveform into 5-minute windows, then estimated univariate Gaussian distributions for each window to approximate the true data distributions, resulting in $\{ [\mu_{i1}, \cdots, \mu_{i7}], \nu_i \}_{i=1,\cdots,3533}$ samples (as presented in Figure~\ref{fig:case-study-demo}).
We then evaluate the Gauss-DIDO regression in 100 replications, where 3533 samples were splitted into 70\% training set and 30\% testing set in each replicate.  

\begin{figure}[h!]
    \centering
    \includegraphics[width=0.65\textwidth]{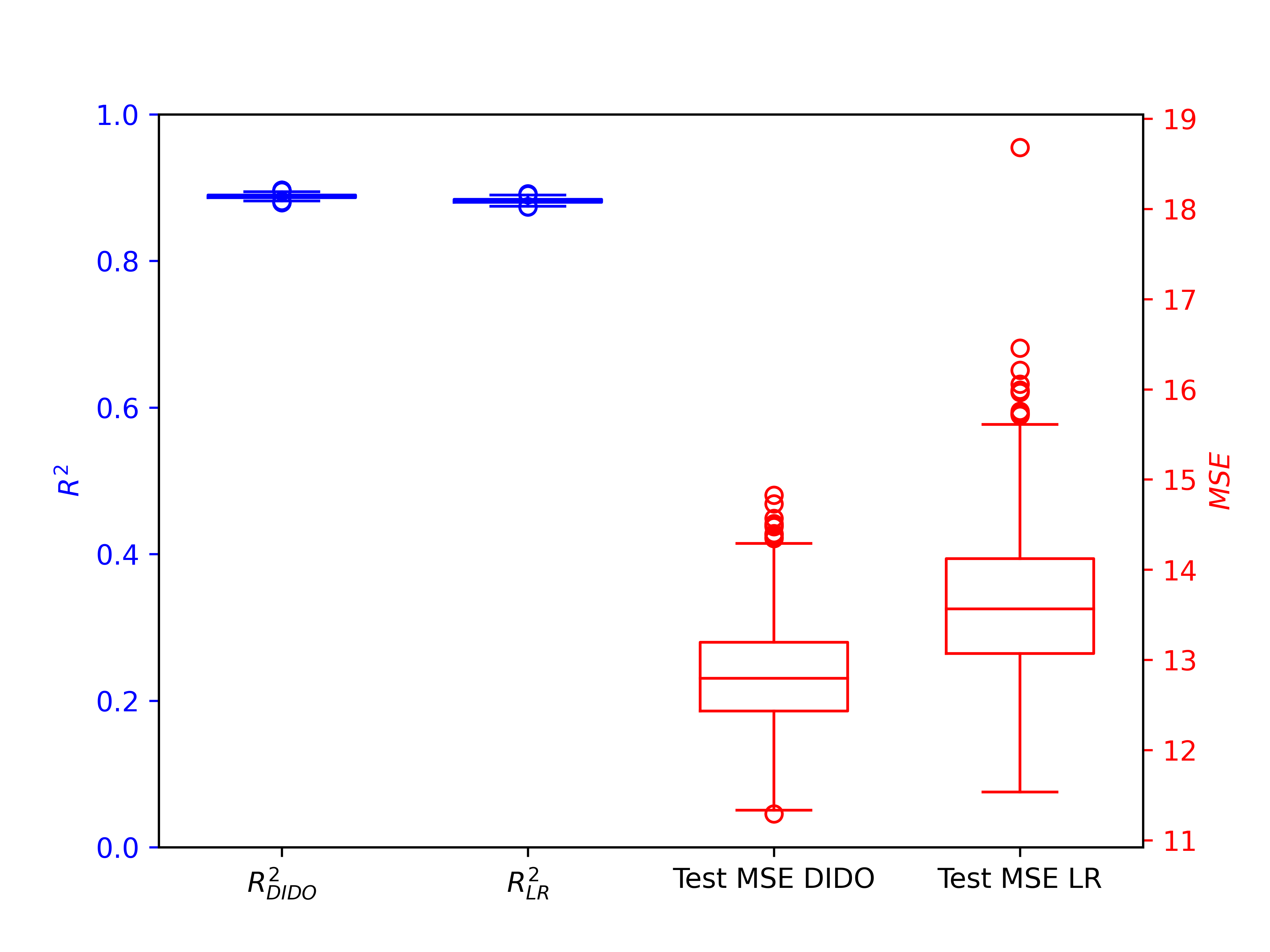}
    \caption{Performance of Gauss-DIDO in a intraoperative cardiac output prediction case study.}
    \label{fig:case-study}
\end{figure}

\begin{figure}[h!]
    \centering
    \includegraphics[width=\textwidth]{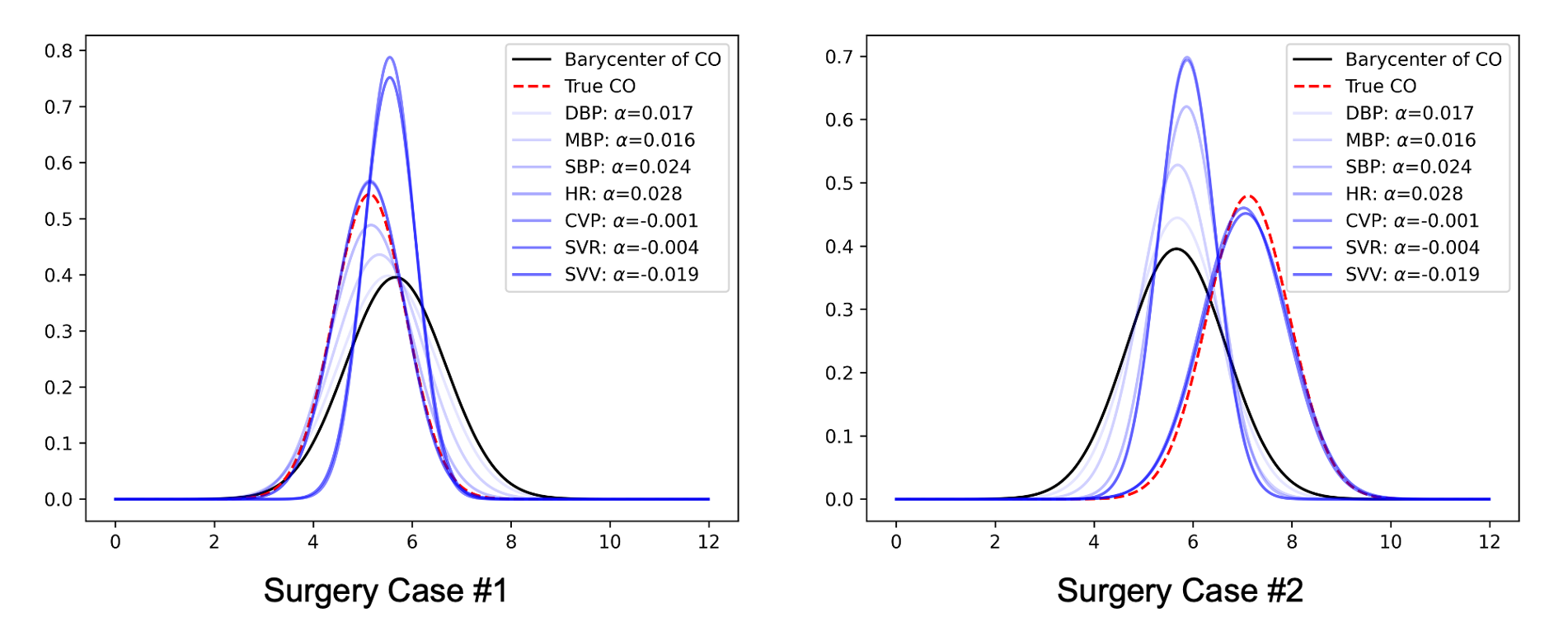}
    \caption{Sequential transport trajectory of two surgery cases.}
    \label{fig:pushforward}
\end{figure}

We summarize the performance in a box plot as shown in Figure~\ref{fig:case-study}. It can be concluded that the DIDO model can significantly better fit the predictors-response relationships (i.e., greater $R^2$); and the DIDO model achieves significantly lower test MSE than LR in predicting the CO distributions.
In addition to the summary statistics, in Figure~\ref{fig:pushforward} we also visualize the sequential transport trajectory from $\bar{\nu}_i$ to the prediction $\hat{\nu}_i$. Here, the black line represents the barycenter of CO $\bar{\nu}_i$, red dotted line represent the true CO ${\nu}_i$, and transparent blue-colored lines represent the sequential transport trajectory based on the sequentially added predictor variables,   where we use lighter to darker transparency identifies the sequence transport. 
It can be observed that each predictor contributes to the transport differently. 
The absolute values of estimated model coefficients are all less than 1.0, due to the fact that the support of all predictors are much larger than the support of the response, identifying the needs to ``standardize'' the predictors. 

\section{Summary}
\label{sec:conc}
Regression analysis with probability distribution predictors and probability distribution response has been an important direction to elevate the current statistical modeling methods to more general and practical scenarios when predictors and responses are sampled from random processes. 
Defining an distribution-in-distribution-out regression model requires one to place the predictors and responses in a metric space, i.e., the Wasserstein space. However, as a Riemannian manifold, addition and scalar multiplication of probability measures are not well-defined due to the non-flat geometry, which hinders the development of DIDO regression model.
In this work, we introduce parallel transport to define multiple linear regression in 2-Wasserstein space by defining the arithmetic operations and prove the additivity. 
The close-form solution of the Fr\'echet least squares estimator is found and proven to be the best linear unbiased estimate. 
To accommodate practical needs, we also derive the special case of DIDO regression model when predictors and response are all Gaussian distributions, resulting in the Gauss-DIDO regression model, which is then evaluated in both simulation study and a real case study.
We also show the connection from Gauss-DIDO regression model to linear regression model in Euclidean space.

The proposed DIDO regression model can open many topics for further research. For example, accelerating the estimation process of the DIDO regression model with free-shaped distributions remains a challenge. Besides, enabled by Corollary~\ref{cor:quad-estimator}, one can easily create some regularized DIDO regression methods to find significant variables. The way to understand the selection of important distributions is unclear. Besides, existing design and analysis of experiments are generally considering scalar factors (\citealt{wu2011experiments, xiao2024modeling, kang_deng_jin_2023}), DIDO is expected to open new opportunities for the investigation of optimal design theories with distributional factors. 
The proposed arithmetic operators can also be extended to sparse estimation of Fr\'echet variance-covariance matrix for measure-valued data (\citealt{fang2016bayesian}).
In practice, the Gauss-DIDO model will enable one to find the relationship among probability distributions, which will enable more efficient knowledge discoveries.

\bibliographystyle{agsm}

\bibliography{Bibliography-MM-MC}
\end{document}